\documentclass[10pt, prd, amssymb,aps,twocolumn]{revtex4}
\tighten
\begin{document}
\draft

\title{Patterns of Symmetry Breaking in Cosmology and the
Laboratory}

\author{R.\ J.\
Rivers \thanks{r.rivers@ic.ac.uk}}

\address{{\it
Theoretical Physics Group; Blackett Laboratory, Imperial College,
London SW7 2AZ}}

\begin{abstract}

In these lectures we look for parallels between symmetry breaking
in the early universe and condensed matter systems, and discuss
experiments that display these.

\end{abstract}
\maketitle

\section{Overview}

Our understanding of the early universe is strongly hampered by
the difficulty in making clean inferences about the nature of its
underlying field theory from the observational data.
 The goal of  the COSLAB
 programme is
 to find general physical principles common to particle
 astrophysics (Cosmology) and condensed matter physics (the Laboratory)
 which permit direct experimental testing in the latter that is impossible in the former.

Phase
 transitions are ideal in this regard, occurring in both condensed matter systems and the
 early universe and, in the former case, being directly testable in the laboratory.
In these talks I shall examine some of the parallels that phase
changes suggest, their consequences and their experimental
 confirmation.
 Because of the breadth of these talks, in many cases I have cited
 review articles, rather than the original references. Brevity requires occasional over-simplification.
 I apologise to authors who feel unfairly omitted.
 In
 particular, I refer the reader to the proceedings of earlier
 COSLAB meetings \cite{COSLAB1,COSLAB2}.



\section{Phase transitions}

Phase transitions are changes of state, associated with changes of
(internal)
 symmetries of the system. Although they occur in both the early
 universe and condensed matter systems, their status is different
 in the two cases:

 In the case of the early universe, we have
many transitions anticipated on general grounds. However, although
we have little knowledge of the symmetries at very high energy
scales, we can make some generic
  predictions. For example, simple models typically produce monopoles and cosmic strings, both of which are observable.
Unfortunately our understanding of the non-equilibrium dynamics of
the very early universe that would enable us to predict the number
of each is very poor (but see Paul Shellard's lectures in these
proceedings).

On the other hand, for condensed matter, many transitions are
known, for which
 our detailed knowledge is very good.
However, there has been limited experimentation on them in
directions parallel to the interests of the cosmologists, which
emphasise their non-equilibrium nature. For these reasons, phase
transitions in condensed matter provide an ideal testing ground to
see to what extent one community can inform the other.

In condensed matter systems familiar changes of state include:
\newpage
\begin{itemize}
 \item Non-Magnetic $\rightleftarrows$ Magnetic
 \item Normal Fluid$\rightleftarrows$ Superfluid
 \item Normal Conductance $\rightleftarrows$ Superconductance
 \item Bose-Einstein Condensation
 \item Liquid crystal formation
 \end{itemize}
There is no hierarchy within these transitions.
In particular, they are true transitions, rather than rapid, but
smooth, cross-overs. The situation is very different for the early
universe, for which early times imply higher energy which, in
turn, implies
 greater symmetry. In consequence,
a cooling universe generates a hierarchy of
 transitions:
\\
\\
\begin{tabular}{ll}
 $\bullet$ Planck Scale &$k_BT\sim 10^{19}GeV\,\,
 (10^{-45}s)$\\
 &\\
 $\bullet$ GUT Transitions $^{**}$ & $k_BT\sim 10^{15}GeV\,\,
 (10^{-37}s)$\\
 &\\
 $\bullet$ Supersymmetry Breaking $^{*}$ & $k_BT\sim 10^{3}GeV\,\,
 (10^{-12}s)$\\
 &\\
 $\bullet$ Electroweak Transition $^{***}$ & $k_BT\sim 10^{2}GeV\,\,
 (10^{-10}s)$\\
 &\\
 $\bullet$ Quark-Hadron Transition $^{****}$ & $k_BT\sim 10^{2}MeV\,\,
 (10^{-4}s)$
 \end{tabular}
\\
\\
\noindent In practice, greater symmetry implies less knowledge
(the more stars the better). The 4-star hadronic transition is
best known. Its cold phase (the hadrons of accelerator physics) is
extremely well understood. The transition is unique among early
universe transitions in that the hot phase, the quark-gluon plasma
(QGP), can be simulated  experimentally, in some sense, by
colliding heavy ions at high enough energy. Our theoretical
understanding of deconfinement is also good, even if we cannot pin
down all the details. However, there are also high baryon number
superconductor/superfluid transitions that could be important in
neutron stars, which we cannot recreate. As we shall see, the
transitions are often cross-overs, but this will turn out to be
one of the less important differences.

 For the 3-star electroweak transition particle accelerators have given us the main properties of
 the cold phase (but for the Higgs sector). The hot phase has been
  estimated theoretically to suggest a
  complicated transition,
  again a crossover,
  but in this case there is no direct experimental
  confirmation.

 For our 2-star GUT transitions there is no direct experimental evidence on either
 side of the transition temperature(s). Our belief in their
 likelihood is largely motivated by the observed convergence of coupling strengths
 at GUT scales due to the screening behaviour of field
 fluctuations.  Our 1-star supersymmetry transition, which balances fermionic against
bosonic degrees of freedom,
  has even less direct support.
 We are motivated by the solution that supersymmetry brings to
 the hierarchy problem (the need to keep GUT and electroweak symmmetry breaking scales separate)
 and the fact
 that theories that incorporate gravity at a quantum level require
 supersymmetry. At the Planck scale we know nothing reliably,
 but it sets the absolute
 scale against which all others are compared.


\section{Patterns of Symmetry Breaking}


For true transitions a change of phase is typically identified by
the vanishing of an {\it order parameter} (or of one or more
components of an order parameter vector or matrix). We have two
cases to consider:

\begin{itemize}
 \item Bosonic Theories: The order parameter fields are bosonic $\phi_a(x)$
 from which we construct order
 parameters as ground state expectation values ${\bar\phi}_a =\langle\phi_a\rangle$.
 Typically, the change in
 symmetry is implemented by
 symmetry breaking in a scalar (semi)classical field potential.
 \item Fermionic Theories:
 Fermionic fields have zero ground state expectation values.
 The order parameters are constructed from expectations of bosonic fermion bilinears, such as
 $\langle{\bar\psi}\psi\rangle$ (quark
 condensate)
 or
 $\langle\psi^\uparrow\psi^\downarrow\rangle$ (Cooper Pair).
Often we can write the underlying fermionic theory as an effective
bosonic order-parameter field theory, so that we can still use
simple scalar potentials to characterise phases.
\end{itemize}
There are no problems for {\it global} symmetry transformations,
exemplified by $\phi(x)\rightarrow e^{ie\alpha}\phi(x)$, where
$\alpha$ is constant in space-time. However, for {\it local}
symmetries, most simply of the form $\phi(x)\rightarrow
e^{ie\alpha(x)}\phi(x)$, with space-time dependent $\alpha (x)$,
the symmetry requires the existence of gauge fields that undergo
their own transformations

For global symmetries the exemplary groups are
 the Orthogonal Groups $O(N)$, with a vector ${\vec \phi (x)}$ of order parameter fields
$\phi_a(x)$, $a=1,2,...N$, transforming under $O(N)$ as
 $\phi(x)\rightarrow
R\phi(x)$. $R$ is a real $N\times N$ orthogonal matrix, with
${\tilde R} R=I$. If $det R =1$ the symmetry group is $SO(N)$,
rather than $O(N)$. The simplest invariant from which to construct
the potential is ${\vec \phi}^2$. [We should also consider the
additive Groups $Z_N$, for which $\phi(x)\rightarrow
U\phi(x);\,\,\,U^N=I$. The most common of these is $Z_2
:\phi(x)\rightarrow -\phi(x)$ for a single real field.]

For local symmetries the exemplary groups are the  Unitary Groups
$U(N)$, although some global symmetries are of this form (e.g. our
$U(1)$ example above). The order parameter fields most simply
transform under the $N$-dimensional vector representation as
$\phi(x)\rightarrow U(x)\phi(x)$, where $U^{\dag} U=I$.  With the
$N$-component $\phi$ complex, the basic invariant is
$\phi^{\dag}(x)\phi(x)$. If $det U =1$ then the group is $SU(N)$.
As we noted, the requirement of local symmetry invariance demands
the existence of gauge fields $A^a_{\mu}(x)$, which transform
under the adjoint representation of the group.
 If we restrict
ourselves to unitary groups $U=U(x)$ there are $N^2$ gauge fields
 $A^a_{b\mu}(x)$ ($a,b =1,2,...,N$) before gauge fixing. For $SU(N)$, with $A^a_{b\mu}(x)$ traceless,
 there are $N^2-1$ gauge fields.

Note that $U(1)= SO(2):\phi^{\dag}\phi = \phi_1^2 +\phi^2_2$. Also
note that
 $SU(2)\approx SO(3)$,\,\,\,\,\,$SO(3)\times SO(3)\approx SO(4)$,
 where the $\approx$ means identical behaviour for infinitesimal
 transformations, but different large-scale behaviour.
Other groups ($Sp(N), G_2, E_2, ...$) are rarely used.

\subsection{\bf General Symmetry Breaking}

Transitions occur when the system has degenerate ground states.
That is, the symmetry group $G$ of the Hamiltonian, or the Action
(Lagrangian), is not the symmetry group $H$ of the ground states
of the system. We write $G\rightarrow H$. $H$ is determined by
enumerating the ground states $\phi_0$. For all $h\in H \subset
G$, we have $h\phi_0 =
 \phi_0$. This enables us to identify the manifold of ground states  ${\cal M}$ as the
 left cosets ${\cal M} = G/H$.

From our comments above, we shall restrict ourselves to
spontaneous symmetry breaking (SSB) visible in a classical, or
effective, bosonic potential.
However, sometimes the symmetry of the Hamiltonian (and the
degeneracy) is only approximate, and we may need additional
 explicit symmetry breaking
by mass terms, etc. Such terms can induce  significant
quantitative changes.  In particular, true transitions are
replaced by crossovers.
Details on symmetry breaking can be found in many sources,
particularly Kibble in \cite{COSLAB1,COSLAB2} and we can only
recreate the rudiments here.

 \subsection{Some simple examples}

The simplest symmetry breaking is that of the $Z_2$
($\phi\rightarrow -\phi$) symmetry of the generalised Ising model
of a single real field $\phi$, with double-well potential $V(\phi
) = {1\over 4}\lambda (\phi^2 - \eta^2 )^2$, and action
\begin{equation}
S = \int d^4x \bigg[{1\over 2}\partial_{\mu}\phi\partial^{\mu}\phi
- {\lambda\over 4} (\phi ^2 -
 \eta^2)^2],
 \label{Z2S}
\end{equation}
  or free energy
 \begin{equation}
 F = \int d^3x \bigg[{1\over 2}(\nabla\phi )^2 + {\lambda\over 4} (\phi ^2 -
 \eta^2)^2].
 \label{Z2F}
 \end{equation}
 The maximum of $V$ at $\phi = 0$ is unstable, but on expanding
about the minimum $\phi = \eta$, as $\phi = \eta + h$, we see that
$V(\phi ) = {1\over 2}m_{h}^2 h^2$ + ({\it non-quadratic terms})
describes $h$-field excitations of mass $m_{h} =
\eta\sqrt{2\lambda}$. Expanding about $\phi = -\eta$ gives an
identical spectrum.
$Z_2$ symmetries have a natural role in Grand Unified Theories
(GUTs) of the early universe, where they provide a simple
selection rule against the rapid decay of the proton.

The next model, to which we shall return repeatedly, is the
Goldstone model of a complex field  $\phi = \phi_1 + i\phi_2$,
with Mexican-hat/wine-bottle potential $V(\phi )  = {\lambda\over
4}(|\phi |^2 - \eta^2 )^2$ (and kinetic/gradient terms changed
correspondingly in (\ref{Z2S}) and ({\ref{Z2F})). The global
$O(2)$ symmetry of rotations in the complex $\phi$ plane is
totally broken by the inequivalent, but equally acceptable,
degenerate ground states with $|\phi | = \eta$, which form the
manifold ${\cal M}= S^1$ (the 1-sphere, or circle). Expanding
about $\phi = \eta$ say, shows that the particle spectrum is that
of a massive (Higgs) particle, mass $m_h$ as above, corresponding
to radial oscillations in the complex $\phi$-plane, and a massless
(gapless) mode, the Goldstone particle, corresponding to
translations of the field along $S^1$.

The natural generalisation of the Goldstone model is to $N$ scalar
fields $\phi_a$, $a = 1,2,...,N$, with potential $ V={\lambda\over
4}({\vec\phi}^2-\eta^2)^2 ={\lambda\over 4}(\phi_a^2-\eta^2)^2$,
where
  $\phi_a^2 = \phi_1^2 +\phi_2^2 + ...$. The $O(N)$ symmetry is
  broken as $O(N)\rightarrow O(N-1)$, with a manifold of groundstates
  ${\vec\phi}^2 = \eta^2$,
  the $N-1$ dimensional sphere ${\cal M} = S^{N-1}$. If we expand about
  $\phi_1 = \eta, \phi_2 = \phi_3 =... =
  0$, the oscillations describe $1$ massive (Higgs) field in the radial field direction
  and $N-1$ massless (Goldstone) fields along the surface of the
  sphere of groundstates. Such a model is termed the $O(N)$ Linear Sigma Model ($L\sigma
  M$).  The $O(N)$ Non-Linear Sigma Model ($NL\sigma M$) is an extreme
  version of this, describing classical rotors, in which, by fixing $\phi_a^2 = \eta^2$,
  we eliminate the Higgs field as a degree of freedom. There are
  no known Goldstone modes in particle physics, although there are
  very light degrees of freedom (e.g. pions), but Goldstone modes
  arise naturally in condensed matter physics as phonons.

For gauge theories it is not sufficient to just examine the
potential $V(\phi )$, because the local nature of the gauge
transformations enforces additional interactions through the
covariant derivatives that extend the kinetic terms. The simplest
case is the $U(1)\rightarrow 1$  Abelian Higgs model for complex
field $\phi$ and gauge field $A_{\mu}$. The action is

 $$ S = \int d^Dx \bigg[|D_{\mu}\phi |^2 - {\lambda\over 4} (|\phi |^2 -
 \eta^2
 )^2 - {1\over 4}F_{\mu\nu}F^{\mu\nu}\bigg],$$
 \\
 where $D_{\mu} = {\partial\over {\partial x_{\mu}}} -
 ieA_{\mu}$ is the covariant derivative, and $F_{\mu\nu}$ the
 electromagnetic field tensor. Although
 ${\cal M} = S^1$, as for the Goldstone model, there are no Goldstone
 modes.
The rearrangement of the field degrees of freedom gives rise to
one real Higgs particle, and  one {\it massive}
 vector field. In some sense the Goldstone mode is swapped for the transverse degree of freedom of
 the vector field.

\section{Symmetry Breaking in Condensed Matter}

The simple forms of SSB given above are most easily realised in
condensed
 matter theory.
We shall largely restrict ourselves to continuous transitions for
which relevant experiments have been performed.

 \subsection{Superfluid $^4He$:}

 This is the global $O(2)$ Goldstone model already discussed, with
 complex order parameter field $\phi = \rho e^{i\theta}$.
 In the 2-fluid model $n_s = \rho^2$ is the superfluid density and
  ${\bf v}_s =
 \frac{\hbar}{m}\nabla \theta$ the superfluid velocity. The Goldstone mode describes sound.
 As before,
 ${\cal M} = S^1$.

\subsection{Low -$T_c$ Superconductors}

The bosonic effective order parameter field $\phi$ is derived from
the $L=S=0$ Cooper pairs of electrons (with momenta close to the
Fermi surface) as
$\phi\sim\langle\psi_{\downarrow}\psi_{\uparrow}\rangle_{L=S=0}$.
This is an Abelian Higgs model with Free Energy
\begin{equation}
F = \int d^3x\bigg[{\hbar^2\over 2m^*}\bigg|\bigg(-i\nabla -
{e^*\over\hbar c}{\bf A} \bigg)\phi\,\bigg|^2 + \lambda (|\phi |^2
-\eta^2)^2 \bigg],
\end{equation}
in the absence of external fields, where $e^* = 2e$, $m^* = 2m_e$
characterise the Cooper pair. The Meissner effect, whereby the
magnetic field penetrates into the bulk superconductor, has its
penetration length determined by the mass of the vector field in
the broken phase.

\subsection{High-$T_c$ Superconductors}

High-$T_c$ superconductivity is a complicated phenomenon. For our
purposes, we adopt an idealised explicitly broken $SO(5)$ model in
$D=2$ dimensions \cite{zhang}. The basic idea is that doping an
antiferromagnet leads to d-wave superconductivity. The bosonic
effective order parameters are, again, constructed from fermionic
bilinears.

We begin with a global $O(3)_{AF}$ L$\sigma$M for
antiferromagnetism with order parameter field (staggered
magnetism) ${\vec n}$ with potential $V({\vec n}) = {\lambda\over
4}({\vec n}^2 - 1)^2$.
This is extended to a
five-component order parameter field
 ${\vec N}= (\phi_1, n_1, n_2, n_3,\phi_2)$, with $O(5)$-invariant
 potential $V({\vec N}) = {\lambda\over 4}({\vec N}^2 - 1)^2$.
[We can work equivalently \cite{zhang,arovas} with a NL$\sigma$M,
in which $|{\vec N}| = 1$.] Ultimately we shall couple $\phi =
(\phi_1 +i\phi_2)$ locally to the EM field as in Low-$T_c$
superconductors.

In the first instance ${\cal M} = S^4$. We break the $O(5)$
invariance {\it explicitly} to $O(3)_{AF}\times SO(2)$ by the
  addition of antiferromagnet and doping terms to the potential,
  most simply as \cite{alama}
\begin{equation}
V(\phi,{\vec\eta}) = {1\over 2}a_S|\phi|^2 + {1\over 2}a_A{\vec
n}^2 + {b\over 4}(|\phi |^2 + {\vec n}^2)^2, \label{UniV}
\end{equation}
 where $a_S,a_R<0$.
We then couple $\phi$ to EM as
\begin{equation}
F = \int d^2x\bigg[{\hbar^2\over 2m^*}\bigg|\bigg(-i\nabla -
{e^*\over\hbar c}{\bf A} \bigg)\phi\,\bigg|^2 + {\hbar^2\over
2m^*}(\nabla{ {\vec n}})^2 + V(\phi,{\vec n}) \bigg].
\end{equation}
Increasing the doping drives $|a_S|>|a_R|$, making the $U(1)\sim
SO(2)$ superconducting direction the
 global minimum.  That is,
 $O(5)\rightarrow O(3)_{AF}\times SO(2)\rightarrow SO(2) $, with ${\cal M} =
 S^1$. This $U(1)\sim SO(2)$ is then broken as for the low-temperature superconductors.

\subsection{Superfluid $^3He$}

 $^3He$ is a Fermi Liquid, which can become superfluid by the formation of
 p-wave (L=S=1) 'Cooper pairs' of $^3He$ atoms. L and S are uncoupled at short
 distances,
 to give a global symmetry group $G = SO(3)_L\times SO(3)_S\times
 U(1)_N$.

 The effective order parameters \cite{volovik} form a $3\times 3$ matrix $A_{\alpha
 i}(x)$, formed from the Fermi bilinears
$\langle\psi (x)\psi (x)\rangle_{L=S=1}$. The label $i = 1,2,3$ is
the orbital angular momentum label and $a=1,2,3$ the spin label.
The $U(1)_N$ describes the overall phase freedom. Above the
transition all the elements of the matrix have zero values. Below
the transition, some of these quantities become non-zero. The
symmetry of the order parameter after the transition corresponds
to the manifold of symmetries which remain unbroken.

 The
free energy of these states can be expressed in the framework of
the phenomenological Ginzburg-Landau theory by a potential
\cite{Bunkov}
\begin{eqnarray}
V_{GL}(A) &=& -\alpha A^*_{a,i} A_{a,i} + \beta_1 A^*_{a,i}
A^*_{a,i} A_{b,j} A_{b,j}+ \nonumber\\
&&+\beta_2 A^*_{a,i} A_{a,i} A^*_{b,j} A_{b,j} + \beta_3 A^*_{a,i}
A^*_{b,i} A_{a,j} A_{b,j} \nonumber
\\
 &+& \beta_4 A^*_{a,i} A_{b,i} A^*_{b,j} A_{a,j} + \beta_5
A^*_{a,i} A_{b,i} A_{b,j} A^*_{a,j}. \label{VGL}
\end{eqnarray}
The different possible symmetries of the order parameter $A_{ai}$
are identified with local minima and saddle points in this
18-dimensional energy surface. There are two stable phases;
 \begin{itemize}
\item The $A$ phase, in which
\[
SO(3)_S\times SO(3)_L\times U(1)_N\rightarrow SO(2)_{S_z}\times
U(1)_{L_z - N/2}\times Z_2.
\]
The manifold of ground states is ${\cal M}_A = G/H_A = S^2\times
SO(3)/Z_2$. The order parameter in the $A$ phase ground state is
anisotropic in both spin and orbital spaces (the 'axial' state). :
most simply, it takes the form $A^0_{a,i} = \Delta_A\,{\hat
z}_{a}({\hat x}_i + i{\hat y}_i)$, where ${\hat x},{\hat y},{\hat
z}$ are unit vectors in the $x,y,z$ directions respectively.

\item The $B$ phase, in which the orbital and spin angular momenta
are locked together as
\[
SO(3)_S\times SO(3)_L\times U(1)_N\rightarrow SO(3)_{S+L}.
\]
The manifold of ground states is now ${\cal M}_B = G/H_B =
S^1\times SO(3)$. As a result, $A_{a,i}$ resembles a rotation
matrix. Specifically, in the bulk $B$ phase, $A_{a,i}$ reduces to
the arbitrary orthogonal rotation matrix $R_{a,i}$, $A_{a,i} =
\Delta R_{a,i} e^{i\Phi}$.
 \end{itemize}

The energy balance between the $A$ and $B$ phases is determined by
the relation between the parameters $\beta_i$. At zero pressure,
the $B$ phase corresponds to the absolute minimum, while at
pressures above 20 bar there is a temperature range in which the A
phase is preferred.

 \subsection{Other systems}

There are other systems whose transitions are understood well,
which potentially have parallels with the early universe. In
particular, we would cite
\begin{itemize}
\item Uniaxal nematic liquid crystals, for which the order
parameter in the nematic phase is the non-oriented director vector
with ground state manifold $RP^2 = S^2/Z_2$. The transition is
first order \cite{chuang}. However, when considering symmetry
breaking at the interface of an isotropic-nematic transition the
anchoring of the director at the interface forces it to lie on a
cone \cite{digal}, whereby the order parameter space is a circle
$S^1$, corresponding to the familiar $U(1)$ breaking.
 \item
Bose-Einstein condensates, which allow for a great variety of
symmetry breaking if species of atoms are mixed. For example,
consider two-species BEC (two different ultracold atomic gases in
a trap) with order parameter fields $\phi_a(x)$, ($a = 1,2$). The
trap-independent part of the potential can be written as
\cite{zhitnitsky}
\begin{equation}
V=\lambda_1(|\phi_1|^2 - \eta^2_1)^2 + \lambda_2(|\phi_2|^2 -
\eta^2_2)^2 + \beta |\phi_1|^2|\phi_2|^2,\label{UniV2}
\end{equation}
with tunable parameters, which is no more than (\ref{UniV})
rewritten for $SO(4)$. For a single species ($1^{st}$ term) we
have an $O(2)$ symmetry that is totally broken, as in $^4He$, and
for two species an $O(2)\times O(2)$ symmetry, broken to $ O(2)$,
with similarity to the breaking of the residual symmetry in
high-$T_c$ superconductors.
\end{itemize}

 \section{Symmetry Breaking in the Early Universe}

 In general, the overall patterns of symmetry breaking in the field theories that we
 believe describe the early universe are more complicated than those in condensed
 matter. In particular, the transitions about which we are most
 knowledgable (later in time) tend to be crossovers, and the ones we know least
 are likely to be complicated by virtue of their belonging to
 larger symmetry groups.

 \subsection{The Quark-Gluon System of Quantum Chromodynamics (QCD):}

Most known particles are strongly interacting (hadrons), of which
the proton, neutron and pion are the lightest members. They are
built from quarks (fermions) and gluons (vector bosons). Quarks
are described by fields $q_{f,c}$, carrying two labels. The label
$c=1,2,3$ is that of a local $SU(3)_c$ symmetry, termed {\it
colour}. In ordinary hadrons the gauge field gluons $A_{\mu c}^a$
($a,c =1,2,3$), demanded by the local symmetry, form a massless
octet ($8 = 3^2 -1$).The fact that there are are no free quarks in
hadronic matter at zero chemical
  potential and temperature (particle accelerators) is explained
  if all known hadrons are colour {\it singlets}.
The other label $f= 1,2,..,N$ describes a global $SU(N_f)$ flavour
symmetry that is intrinsically broken. At $N_f = 2$ this breaking
is very small (electromagnetic), at $N_f= 3$ it is still fairly
small. For $3<N_f\leq 6$ the breaking is large.
The resulting theory of quarks and gluons is known as Quantum
Chromodynamics. 

The transition from a quark-gluon plasma to hadrons at low
chemical potential (the early universe and heavy-ion collisions)
is complicated by having two different aspects. The first is the
approximate breaking of chiral symmetry that is a consequence of
the lightness of the common quarks, for which the gluons play a
subsidiary role. The second is the confining/deconfining nature of
the gauge sector, for which the quarks are relatively unimportant.
Although QCD can be tackled directly in some circumstances
(through lattice gauge theory), it is informative to think of it
as a modification of one of these extremes. More details can be
found in the review article by Rajagopal \cite{rajagopal}.

\begin{itemize}

 \item
 {\bf Low energy Chiral theory:} Common hadrons are built from two
 quarks (termed {\it up}
and {\it down}) which are approximately massless. One important
idealisation is to take them massless in the first instance, $m_d
= m_u = 0$, to give a theory that is invariant under independent
transformations of the left and right hand components of the quark
doublet $\psi = \psi_L + \psi_R$. Writing these as
 $\psi_L\rightarrow L\psi_L$ and $\psi_R\rightarrow R\psi_R$,
 where $L,R\in SU(2)_f$ we would have degenerate parity doublets
 if the $SU(2)_L\times SU(2)_R$ chiral symmetry remains unbroken.
 This is not the case experimentally. Breaking the symmetry as $SU(2)_L\times SU(2)_R\rightarrow
 SU(2)_{L+R}$ gives pions as the necessary Goldstone bosons.
 The pions are not massless (there are no known Goldstone bosons
 in particle physics) but they are anomalously light, requiring
 explicit (but small) symmetry breaking. This makes the transition
 a crossover.
 In fact, there is an additional $U(1)_A$ that is broken by instantons, without
 Goldstone bosons, that we shall not consider here. Extensions to
 $N_f=3$ (requiring 'strange' quarks with $m_s = 0$) with a
 more badly broken $SU(3)_L\times SU(3)_R\rightarrow SU(3)_{L+R}$
 will be made later.

\item {\bf Effective $O(4)$ sigma models}: Still in the context of
chiral theory, the chiral order parameter, a $2\times 2$ matrix
$M$ in flavour space, can be written in terms  of a pion triplet
${\vec \pi}$, with $\pi^+ = \pi_1+i\pi_2$, etc, and a singlet
$\sigma$, as $\langle{\bar\psi}^a_{L\alpha}\psi^{\alpha
b}_R\rangle = M^{ab} = \sigma\delta^{ab} +
{\vec\pi}.({\vec\tau})^{ab}.$ Low-energy pion physics is well
represented by an effective $O(4)\,\, L\sigma M$ , with potential
$V= \lambda ({\vec\pi }^2 +\sigma^2 -f^2)^2$ (or an $O(4)\,\,
NL\sigma$M with the constraint ${\vec\pi }^2 +\sigma^2 =f^2$) and
explicit term linear in $\sigma$ to enforce $m_{\pi}\neq 0$
mass-breaking. In the absence of this, with manifold ${\cal M} =
S^4$ of ground states the pions are the Goldstone modes (and the
Higgs is the $\sigma$). Note that $O(4)\rightarrow O(3)\approx
O(3)\times O(3)\rightarrow O(3)\approx SU(2)\times
SU(2)\rightarrow SU(2)$.

\item {\bf Deconfinement} If we take quark masses to be infinitely
large, we recover a pure gauge theory. In this case we can
construct  a non-local order parameter (the Polyakov loop), whose
$Z_3$ symmetry-breaking characterises deconfinement
\cite{korthes-althes}. This is complicated by finite quark masses
and is too subtle for us here.
\end{itemize}
We should stress that there are many other ways to tackle the
transitions, but they are even less relevant to our subsequent
discussion.






\begin{itemize}
\item {\bf Colour Superconductivity:} At high baryon density
(chemical potential) there is the possibility of Colour
Superconductivity/Superfluidity. This cannot be created in the
laboratory but possibly exists in neutron stars, which could have
quark matter cores. In cold dense quark matter the relevant
degrees of freedom are those of quarks with momenta near to the
Fermi surface. Quark attractions will lead to the creation of
Cooper pairs. If we begin by assuming unbroken $SU(3)_f$, with
massless quarks $m_u= m_d = m_s=0$, Cooper pairs cannot be flavour
singlets, and both colour and flavour symmetry is broken. The
unbroken symmetry is (where $L$ and $R$ are chiral flavour)
 $$G = SU(3)_L\times SU(3)_R\times SU(3)_c\times U(1)_B,$$
where $U(1)_B$ counts baryons. The possibility exists of
colour-flavour locking (CFL), by condensates
$\langle\psi_L^{\alpha a}\psi_L^{\beta b}\rangle_{\bar 3}\propto
\Delta\epsilon^{\alpha\beta\gamma}\epsilon^{ab\gamma}$, where
$\alpha, a$ denote colour and flavour respectively. This assumes
that the ${\bar 3}$ channel in $3\times 3 = 6 + {\bar 3}$ is the
attractive channel.
 \begin{eqnarray}
 &&SU(3)_L\times SU(3)_R\times SU(3)_c\times
U(1)_B\rightarrow\nonumber\\
&&\rightarrow SU(3)_{L+R+c}=SU(3)_{f+c}.
 \end{eqnarray}
  Order parameters form
a $3\times 3$ complex matrix $\Sigma^a_b$ built from quark
condensates $\langle\psi_L\psi_L\rangle_{{\bar 3}}$ and
$\langle\psi_R\psi_R\rangle_{{\bar 3}}$, as above, expressible as
the nine Goldstone bosons arising from the Chiral symmetry
breaking. There is an effective L$\sigma$M realisation of this
that we shall not pursue.

Explicit symmetry breaking from mass terms is necessary.  There
are several possible phases as we further break the symmetry by
breaking the $SU(3)_f$ quark mass degeneracy. On introducing a
relatively massive strange quark, $m_s\neq m_u = m_d$, the $K^0$
is the pseudo-Goldstone boson, and readily forms a condensate. We
have what is known as CFL $+K^0$, in which
 $SU(3)_{c+f}\rightarrow SU(2)_{I'}\times U(1)_{Y'}\rightarrow
 U(1).$

The simplest scenario \cite{buckley} has a doublet $\Phi$ of $K^0,
K^+$ charged bosons
 $ \Phi = \left( \begin{array}{c}K^+\\
K^0\end{array}\right)$ for which
\begin{equation}
V(\Phi)= \lambda (|\Phi |^2 - \eta^2)^2 \label{UniV3}
\end{equation}
 and ${\cal
M} = S^3$.

 If we include EM effects so that $m_s\neq m_u\neq m_d$ we
 have the further breaking
  $SU(3)_{c+f}\rightarrow U(1)\times U(1)\rightarrow
 U(1).$ The effect of the explicit mass breaking terms is to
 recreate the potential $V$ of (\ref{UniV2}) for $(\phi_1,\phi_2) =
 (K^+,K^0)$.
\end{itemize}

\subsection{Electroweak Transition/ Standard model:} For
systems of low baryon density, we have the unambiguous pattern of
symmetry breaking of the Salam-Weinberg model,
$$SU(3)_c\times SU(2)_I\times
U(1)_Y\rightarrow SU(3)_c\times U(1)_Q,$$ where $Q = I_3 + {1\over
2}Y$ is the electromagnetic charge. This is replicated for each of
the three families of quarks and leptons (corresponding to $N_f =
(1,2), (3,4), (5,6))$ respectively. Considering only the leptonic
sector, the simplest scenario has a doublet $\Phi$ of charged
bosons
 $$ \Phi = \left( \begin{array}{c}\phi^+\\
\phi^0\end{array}\right)\hspace{1.0cm}\mbox {for which}\,\,
V(\Phi)= \lambda (|\Phi |^2 - \eta^2)^2$$ of (\ref{UniV3}), after
field relabelling. However, in this case $V(\Phi)$ is exact, with
no further explicit symmetry breaking.

On symmetry breaking we have the familiar three massive vectors
($W^{\pm}, Z^0$) and one massless gauge field (the photon). The
spin-1 nature of the massive vectors has been bought at the
expense of three of the field degrees of freedom of the $\Phi$
doublet, to leave one massive real Higgs boson. For  the assumed
Higgs mass ($m_h> 80GeV$) the phase diagram  shows a crossover.

 \subsection{Grand Unified Theories:}

 Ignoring supersymmetry, we assume the existence of symmetries
 $G$ such that
  $$G\rightarrow G_{SM} = SU(3)_c\times SU(2)_I\times U(1)_Y\rightarrow SU(3)_c\times
 U(1)_Q.$$
e.g  $E_6\rightarrow SO(10)\rightarrow SU(5)\rightarrow
SU(3)_c\times SU(2)_I\times
 U(1)_Y.$
In practice, we need an additional discrete symmetry to prevent
too rapid a decay of the proton into leptons that is otherwise
permitted by standard GUT symmetries. Groups like $SO(10)$ contain
a $U(1)_{B-L}$ gauge symmetry, which breaks to give an extended
symmetry $G_{SM}\times Z_2$, forbidding fast proton decay.
Motivated by the work of Witten \cite{witten} and others on
superconducting strings , groups $G$ containing $U(1)\times U(1)$,
with potentials (\ref{UniV}) and (\ref{UniV2}) have been found
useful, for reasons that we shall see later.

 However, once we attempt to accommodate supersymmetry at both a
 minimal level and beyond, the number of possibilities
 proliferates \cite{mairi}.

\section{Signatures of Transitions}

The similarity between transitions in condensed matter and the
early universe is interesting, but for it to be compelling we need
a means of monitoring how transitions take place in condensed
matter, so that we can check if similar assumptions about the
early universe lead to our understanding it better. The way we do
this is by looking for signatures of transitions from which we can
infer the nature of the symmetry breaking.

Consider the simplest case of a double-well potential for a real
scalar field $\phi(x)$ We rewrite the potential $V(\phi) =
{\lambda\over 4}(\phi^2 - \eta^2)^2$ as
$$V(\phi) = {1\over 2}\phi^2 m^2 + {\lambda\over 4}\phi^4 +
constant.$$ $m^2 = -\lambda\eta^2$ is negative. All this is at
zero temperature. As temperature increases and the field becomes a
plasma $m^2$ becomes temperature dependent, increasing towards
zero  as $m^2(T) = m^2(1-T^2/T_c^2)$ (in the simplest
Ginzburg-Landau, or one-loop, approximation). The order parameter
${\bar \phi} = \langle\langle\phi (x)\rangle\rangle$, where the
double average is both thermodynamic and quantum mechanical,
satisfies ${\bar\phi}^2 = \eta^2(T) = -m^2(T)/\lambda$. At the
critical temperature $T_c$ both ${\bar\phi}$ and $m^2(T)$ vanish.
In relativistic quantum field theory $T_c\sim\eta$, the symmetry
breaking scale. For $T>T_c$, $m^2(T)>0$ and the $Z_2$ symmetry is
restored.

Let us now reverse the order, as in the early universe, and cool
the field through $T_c$ from above, breaking the $Z_2$ symmetry.
How is it that symmetry is broken, given the $\phi\leftrightarrow
-\phi$ symmetry of the potential? The standard way, in field
theory textbooks, is to bias the potential by introducing a
spatially uniform external source $j$ coupled to the field as
$j\phi$. For $j\neq 0$ there is now a unique ground state to which
the system relaxes. On taking $j\rightarrow \pm 0$ the system
stays in this groundstate, with ${\bar\phi} = \pm\eta$ according
as the source is removed.

In the early universe there is no such uniform bias. Instead, as
we pass through $T_c$ there are {\it local} thermal fluctuations
of the field that will drive it from the  unstable maximum at
$\phi = 0$ towards one or other of the minima. That is, for some
${\vec x}$, $\phi ({\vec x})$ is driven to $+\eta$, for some it is
driven to $-\eta$, and domains form in which $\phi$ is correlated.
In fact, since the transition is implemented in a finite time,
causality (the finite speed at which the field can order itself)
requires this domain structure. Once the system is behaving
classically, the boundaries of these domains are 'walls', in which
the field passes from $\phi = -\eta$ to $\phi = \eta$ according to
the classical equation $\delta S/\delta\phi= 0$. Such a 'domain
wall' in the $x-y$ plane, with profile $\phi(z) = \eta\tanh
(|m|z/\sqrt{2})$, has thickness $\xi_0\sim |m|^{-1}$ and energy
per unit area (surface tension) $\sigma \sim |m|\eta^2$. We will
know that the transition has taken place by the existence of these
walls. Already we have the powerful result that, if we do not
adopt an inflationary scenario to dilute wall formation, then
domain walls produced at the GUT scale are so energetic that they
would close the universe, in contradiction to the evidence.
Although a closed domain wall can contract and disappear, it is
not possible to remove part of a wall, and a single wall requires
an infinite field reordering to remove. Domain walls are the
simplest entities to show this topological stability, and they
provide the simplest example of a {\it topological defect}.

More generally, as we shall see, topological defects are twists,
or knots, or other deformation of the fields that carry a
topological charge that prevents them unwinding, or untwisting
(but not annihilating). Whether defects exist depends on the
detailed nature of the symmetry breaking. If they do exist,
individual defects produced in the early universe can survive into
the present era as 'fossils'. Observing them would provide
concrete evidence for transitions having taken place and, if they
could be studied, provide information on the symmetry breaking. We
can check this out directly with condensed matter systems that
permit defects. Further, we can make predictions for defect
formation that can be checked in condensed matter, and which have
implications for the early universe.

Of course, for different patterns of symmetry breaking the
frustration in the order parameter fields as they link domains
formed after the transition may not be resolved with topological
defects. In such cases the twists in the fields can unwind at no
great energy cost and, at times well after the transition, there
is no direct signal for it having taken place. In practice, models
for the early universe almost inevitably produce defects of one
form or another, and the problem is more one of not seeing them.

\section{Topological Defects}

\subsection{Vortices and Strings}

The most common topological defect in condensed matter physics and
the early universe is the string, or vortex, seen most simply in
the breaking of a $U(1)$ symmetry.

Consider a global theory of a complex scalar field $\phi (x) =
\rho (x) e^{i\theta (x)}$, with effective potential $V(\phi, T) =
{\lambda\over 4}(|\phi |^2 - \eta^2(T))^2$, where $\eta(T)$,
vanishing at the critical temperature $T_c$, is given above. As we
cool through $T_c$  the field, originally concentrated around the
(then) stable minimum at $\phi = 0$, begins to explore the
possible groundstates with $\rho = \eta$. With no bias to make
$\phi$ fluctuate away from the unstable maximum at $\phi = 0$ with
any particular phase, $\theta ({\vec x})$ will vary from point to
point, subject to continuity.

Let us take a closed path ${\cal C}$ in space. Once the transition
has been completed, $|\phi(x)|\approx\eta$ for $x\in {\cal C}$. If
we make a complete circuit in ${\cal C}$ then $\phi (x)$ executes
a closed path in $S^1$, the circle of ground states. Continuity of
$\phi (x)$ requires that the change $\Delta\theta$ in $\theta (x)$
along the path is an integer multiple of $2\pi$, $\Delta\theta =
2\pi n$.  We can find classical solutions of infinite extension
along the z-axis of the form $\phi_n({\vec x}) = \rho
(r)e^{in\theta }$, where $r^2 = x^2 + y^2$, and $\theta =
\tan^{-1}(x/y)$ is the azimuthal angle, that demonstrate this. In
$^4He$ they are vortices with quantised vorticity, of winding
number $n$.  It happens that, if $n>1$, it is energetically
advantageous for the vortex to break into $n$ vortices of winding
number unity. These are the stable topological defects of the
model, being unbreakable. If we shrink the loop ${\cal C}$ to the
origin $r=0$, we must have $\rho_n(0) = 0$ for $\phi (x)$ to be
defined. That is, the core of the vortex is the false groundstate,
just as the core of domain wall was. In low-$T_c$ superconductors
the coupling of the electromagnetic and Cooper pair fields makes
them Abrikosov vortices with quantised magnetic flex. As happens
for $^4He$, the Cooper pair field $\phi = 0$ at the vortex core.
The thickness of the false vacuum ($\phi\approx 0$) core is the
London length, the thickness of the flux tube the Meissner length.
In the early
universe, with no Goldstone bosons, in general vortices will be of
the Abrikosov type.

Suppose we have collection of vortices of the type above. Consider
two closed paths ${\cal C}_1$ and ${\cal C}_2$ with a point in
common, in which the field takes winding numbers $n_1$ and $n_2$.
We can trivially combine them into a closed path ${\cal C}_1 +
{\cal C}_2$ with winding number $n_1 + n_2$. That is, the mappings
from closed paths ${\cal C}_i$, which we can take to be circles
$S^1$ into the set of ground states ${\cal M} =S^1$ form the
additive group of integers $Z$. The group $Z$ is termed the {\it
Fundamental Group} or the {\it First Homotopy Group} of $S^1$,
written $\Pi_1 (S^1)$.

More generally, a transition will produce topologically stable
vortices if $\Pi_1 ({\cal M}) = \Pi_1(G/H)$ is non-trivial. For
the condensed matter systems that we have discussed above, we also
find topological $Z$ vortices in single-species BEC and liquid
crystals at the interface of an isotropic-nematic transition.  For
our broken $SO(5)$ high-$T_c$ superconductor the situation is more
complex, although vortices have winding number $n\in Z$. As we get
close to critical doping, and $SO(5)$ symmetry is restored (prior
to coupling to electromagnetism) the stable vortices have
non-trivial antiferromagnetic cores in which the order parameter
${\vec\eta}\neq{\vec 0}$, although $|\phi|$ necessarily vanishes
there. As we achieve critical doping the antiferromagnetic core
expands to destroy the vortex, since $\Pi_1(S^4)$ is trivial, not
permitting topological charge.

For superfluid $^3He$ the situation is even more complicated. In
the $A$-phase, $\Pi_1(S^2\times SO(3)/Z_2) = Z_4$, addition modulo
$4$. There are four types of vortices, with winding numbers
$n=0,{\pm}1/2,{\pm}1$ (mod $2$). Depending on the parameter
values, it is again possible to have vortices with non-trivial
cores. In the $B$-phase, $\Pi_1(S^1\times SO(3)) = Z\times Z_2$,
and there are two types of vortices with arbitrary winding number.
Not all vortices are normal (with all components of $A_{\alpha,i}$
vanishing at their cores). It is possible for $B$-phase vortices
to have $A$-phase cores.

For cosmology, topological strings (vortices) appear in colour
superconductivity and superfluidity. In supersymmetric GUT
transitions which solve the monopole problem, lead to baryogenesis
after inflation and constrain proton decay, the vast majority of
models lead to topological (cosmic) strings \cite{mairi}.

We should not ignore so-called {\it embedded} strings, in which,
when $G\rightarrow H$, there are subgroups $G',H'$ of $G,H$
respectively, for which $\Pi_1(G'/H')$ is not trivial. Such
strings occur in the $O(4)$ sigma model of low-energy pions
and in electroweak breaking $SU(2)\times U(1)\rightarrow U(1)$.
However, such strings have no topological winding number and
whether they are energetically classically unstable or not depends
upon the values of the parameters (masses, coupling constants).
Even if they are the role of quantum fluctuations is not clear. In
the important case of electroweak breaking the strings are
unstable to begin with, and can only have transitory effects.

Loops of simple relativistic strings contract and turn their
energy into particle production. Although vortex loops in $^4He$
can be stabilised by Magnus forces, in general loops conract.
However, many theories (most simply of the $U(1)\times
U(1)\rightarrow U(1)$ type) can produce strings or vortices that
have non-trivial cores with angular momentum, which can stabilize
loops (vortons) \cite{zhitnitsky2}.  Such a mechanism is not
possible in $^3He$ on energetic grounds, although it is possible
in 2-component BEC, where it is possible that an example has been
seen. The main case of theoretical interest is that of CFL+$K^0$
quark matter, because of its relevance to neutron stars.

\subsection{Monopoles and more}

Consider an $SO(3)$ sigma model with  3-vector field ${\vec\phi}
(x)$ and potential $V({\vec\phi}) = \lambda/4({\vec\phi}^2 -
\eta^2(T))^2$ as it cools through its critical temperature at
which $\eta(T_c) = 0$ in a finite time. Causality requires the
direction of ${\vec\phi}$ to be uncorrelated at large distances.
This frustration is resolved by classical 'knots' in the field, of
the form ${\vec\phi}(x) = \rho (r){\vec x}/r$, where $r$ is the
radial distance from the centre of the knot. These are the
topological monopoles of the model and cannot be 'unwound'. If we
enclose one by a sphere $S^2$, we see that they are non-trivial
mappings of $S^2$ onto the manifold of ground states ${\cal M} =
S^2$. These mappings again form a group, the {\it Second Homotopy}
group $\Pi_2({\cal M})$. In fact, $\Pi_2(S^2) = Z$, and the
monopole solutions are characterised by integer winding number,
for which our example ($|n|=1$) is stable.

More generally, we have monopoles whenever $\Pi_2({\cal M})$ is
non-trivial. In condensed matter physics we find monopoles in the
$A$-phase of superfluid $^3He$, for which $\Pi_2(S^2\times
SO(3)/Z_2) = Z$.

Monopoles arise inevitably in GUT transitions.  This is a
consequence of the fact that, if $G$ is connected and semisimple
(i.e. $\Pi_1(G) = 1$), then $\Pi_n(G/H) = \Pi_{n-1}(H)$. Thus, if
$\Pi_1(H)\neq 1$, we have monopoles. Since $H$ necessarily
contains $U(1)_Q$, $\Pi_1(H)\neq 1$. Just as domain walls produced
in a non-inflationary universe would close it, so would monopoles.
It was to dilute the monopole density to an acceptable level that
Guth originally introduced inflation. This does not mean that all
defects are diluted.  There is no difficulty in developing hybrid
models in which monopoles are diluted but strings survive
\cite{mairi}.

All the above is only the simplest possibilities. With suitable
symmetry breaking we can have strings ending in monopoles, domain
walls bounded by strings, and so on.

\subsection{The observation of defects}

Let us postpone estimating how many defects we might see at a
transition until later, and ask the simpler question as to whether
we see examples of these defects in real life. All the simple
defects permissable in condensed matter systems are seen in the
laboratory, as are many more complicated combinations of defects.
See articles in \cite{COSLAB1,COSLAB2}

In cosmology the situation is the converse, with no unambiguous
sightings of monopoles or strings (vortices). Cosmic strings can
be best observed by gravitational lensing, in which no central
object is seen. The metric around a cosmic string is conical, with
a defect angle $\delta=8\pi (T_c/M_{Pl})^2\sim 10^{-5}$. Recently
a plausible galaxy string lensing candidate was observed
\cite{sazhin}, but in the absence of further lensing nearby, the
existence of a cosmic string is in doubt. Cosmic strings are, in
principle, a potential source of Extra High Energy Cosmic Rays
(EHECR) \cite{bhattachargee}. However, simple mechanisms like loop
decay, cusp radiation and string interconnection give a flux many
orders of magnitude ($>10$) too small. There are many other
variants, among which vorton loops and loops of superconducting
string are the more promising \cite{bonazzola}. In the latter
case, the superconducting current increases until the loop
disintegrates, producing EHECR.

The most promising source of quark-hadron transitions is the
interior of neutron stars \cite{rajagopal,boyanovsky2}. The
quark-gluon condensate is highly compressible, compared to nuclear
matter. Arguably this is seen in millisecond pulsars. As they
spin-up via accretion the core-density decreases and the core
turns to hadrons. As the quark-matter spins it inhibits an
increase in the pulsar frequency $\omega$ because of an increase
in the moment of inertia. Detailed calculation suggests a
frequency ($\omega\sim 300Hz$) at which the pulsar distribution
should peak, and this is observed experimentally
\cite{glendenning}.

The vortices and vortons that are produced at the transition
provide candidates for the observed glitches in pulsar frequencies
($\delta\omega /\omega\leq 10^{-6}$). The suggestion is that these
are understood as a consequence of the interactions between the
rigid crust of the neutron star and vortices/vortons in the
neutron superfluid interior as they try to move outwards.

\section{The Dynamics of Defect Production}

\subsection{Causal Bounds}

Because phase transitions take place in a finite time, causality
guarantees that correlation lengths remain finite. If the symmetry
breaking permits, defects arise so as to mediate the correlated
regions with different ground states. They provide an excellent
experimental signature for the way in which transitions are
implemented.

For condensed matter systems, Zurek\cite {zurek1,zurek2} suggested
that causality alone is sufficient to determine the initial
density of defects arising in a continuous transition. In this he
paralleled proposals made by Kibble\cite{kibble1} in the context
of quantum field theory models of the early universe.

Consider a system with critical temperature $T_{c}$, cooled
through that
temperature so that, if $T(t)$ is the temperature at time $t$, then $%
T(0)=T_{c}$. ${\dot{T}}(0)=T_{c}/\tau _{Q}$ defines the quench
time $\tau _{Q}$. One formulation (see \cite{zurek2}) of the
Zurek/Kibble causal bound is the following: Suppose, at time $t$,
that ${c}(t)={c}(T(t))$ is the maximum speed at which the order
parameter can respond to a change in the environment. Then, at
time $t$ the causal horizon has diameter
\begin{equation}
h(t)=2\int_{0}^{t}ds\,{c}(s).  \label{causal}
\end{equation}
Let $\xi _{ad}(t)=\xi _{ad}(T(t))$ be the adiabatic healing length
(correlation length), diverging at time $t=0$ ($T=T_{c}$). The
first time that defects can appear is at time ${\bar{t}}$, when
the causal horizon becomes big enough to enclose one of them, $\xi
_{ad}({\bar{t}})=h({\bar{t}}) $.

As a direct consequence, ${\bar{t}}$ has the form ${\bar{t}}=\tau
_{Q}^{1-\gamma }\tau _{0}^{\gamma }$. The scaling exponent $%
\gamma $ depends upon the system. For dissipative condensed
matter, with critical slowing down ($c(0) = 0$), $\tau _{0}\ll
\tau _{Q}$ is the relaxation time of the longest wavelength modes,
for QFT (with $c$ a constant) the time for light to traverse a
Compton wavelength. As a result, $\tau _{Q}\gg {\bar{t}}\gg
\tau _{0}$. If $\xi _{ad}(t)\sim \xi _{0}(t/\tau _{Q})^{-\nu }$ for $t\sim 0$%
, where $\xi _{0}=O(\xi _{ad}(T=0))$, then the initial domain size
and defect separation is predicted to be
\begin{equation}
{\bar{\xi}}\sim \xi _{ad}({\bar{t}})=\xi _{0}\bigg(\frac{\tau _{Q}}{\tau _{0}%
}\bigg)^{\sigma }\gg \xi _{0},  \label{xibar}
\end{equation}
where $\sigma =\gamma \nu $. This is very large on the scale of
cold defects. We term $\sigma $ the Zurek-Kibble (ZK)
characteristic index. In the mean-field approximation $\sigma =
1/4$ typically for condensed matter, and $\sigma = 1/3$ for QFT.

We see that the same arguments apply to crossovers
\cite{bettencourt}, provided they are weak enough that the
correlation length is larger than the causal horizon at time
${\bar t}$, even though it does not diverge. We assume this to be
the case.

\subsection{Unstable modes}
In the several years since these simple bounds were first proposed
we have acquired a much better understanding of the way in which
transitions occur. These does not mean that these bounds have lost
their relevance, but that they need to be qualified.

Any dynamical equations for the onset of a continuous transition
will embody causality, by definition. However, the transition
cannot be said to have happened before the order parameter has
achieved its equilibrium value $|\phi |^2 = \eta^2$, (in our
notation for potentials). If $\langle ...\rangle_t$ denotes
ensemble averaging at time $t$ then a lower bound on the first
time from which we can start counting defects is $t=t^*$, for
which $\langle |\phi |^2\rangle_t =\eta^2$.

For many condensed matter systems and for quantum fields, the way
in which $\langle |\phi |^2\rangle_t$ builds up to its final value
is by the growth of the amplitudes of the unstable long-wavelength
modes, which are unstable  because of the upturned parabolic free
energy density at initial times.
The time $t^*$ is,
crudely, the time for these modes to roll from the top of the hill
to the groundstates at the bottom, related to the spinodal time.

The equations that control this ordering through instabilities
differ for different systems.

\subsection{\bf Dissipative systems:} In the vicinity of $T_c$,
dissipative systems are often well approximated by the
time-dependent Ginzburg-Landau equations (TDGL). For a global
theory these are Langevin equations of the form
\begin{equation}
\frac{1}{\Gamma}\frac{\partial\phi_{a}}{\partial t} =
-\frac{\delta F}{\delta\phi_{a}} + \eta_{a}, \label{tdlg}
\end{equation}
where $\eta_{a}$ is Gaussian thermal noise, satisfying
$\langle\eta_{a} ({\bf x},t)\eta_{b} ({\bf y},t')\rangle =
2\delta_{ab}T(t)\Gamma\delta ({\bf x}-{\bf y})\delta (t -t').$
This is a crude approximation for $^{4}He$, and a simplified form
of a realistic description of $^{3}He$ \cite{Bunkov}, and an
important part of the description of low-$T_c$ superconductors.

This seems a very different picture from that
originally proposed by
Zurek. A priori, $t^*$ is not related directly to the causal
${\bar t}$, but it is not difficult to see why they might be
comparable. Unstable modes grow exponentially fast. As long as
dimensional analysis makes ${\bar t}$ the natural unit in which to
measure time, any exponentially growing term will achieve values
that are not exponentially large at times $t = O({\bar t})$.
This can be checked explicitly for fast quenches on retaining only
the quadratic part of $F$, assuming that growth in $\langle |\phi
|^2\rangle_t$ is then checked by the (linearised) back-reaction.
This is justified by seeing that the
growth in long wavelength modes is largely completed in the linear
regime, when the field components $\phi_a$ are effectively
independent \cite{RKK,rayulti1}.

\subsection{\bf Relativistic QFT:} For QFT the situation is
rather different. In the previous section, instead of working with
the TDLG equation, we could have worked with the equivalent
Fokker-Planck equation for the probability $p_{t}[\Phi ]$ that, at
time $t>0$, the measurement of $\phi$ will give the function $\Phi
({\bf x})$. When solving the dynamical equations for a hot quantum
field  it is convenient to work with probabilities from the start.

The probability $p_{t}[\Phi]$ that, at time $t$, the measurement
of $\phi$ will give the value $\Phi$ is $p_{t}[\Phi] = |\Psi|^2$,
where $\Psi_{0}$ is the state-functional with the specified
initial condition (e.g. Boltzmann distributed).  In the language
of path integrals, $p_{t}[\Phi]$ can be written as an  integral
for fields with increasing time ($\Psi$) followed by an integral
for fields with decreasing time ($\Psi^*$). The time contours can
be joined to give a closed time-path and non-equilibrium
calculations in QFT are termed closed time-path calculations.

In practice, there is no need to calculate $p_{t}[\Phi]$ directly.
If $\langle |\Phi ({\vec x})|^2\rangle_t$ measures the growth of
field modes as an ensemble average with respect to $p_t[\Phi]$,
then
$$\langle |\Phi ({\vec x})|^2\rangle_t = \langle |\phi ({\vec x},t)|^2\rangle$$
for Wightman fields $\phi ({\vec x},t)$, subject to the thermal
boundary conditions at the initial time before the transition is
implemented.  Mode analysis is all but impossible outside
self-consistent linearisation of the back-reaction
\cite{boyanovsky}. Within the self-consistent linear regime (which
is the best that can be done numerically) the mode equations that
determine $\langle |\phi |^2\rangle_t$ are now the classical
second-order equations
\begin{equation}
\frac{\partial^{2}\phi_a}{\partial t^{2}} = -\frac{\delta
F}{\delta\phi_a}. \label{op}
\end{equation}

Yet again, this looks a rather different picture from that
originally proposed by Kibble, in which the system froze in before
the transition was effected. Nonetheless, although $t^*$ is not
related directly to the causal ${\bar t}$, they are again
comparable, essentially because unstable modes grow exponentially
fast, and ${\bar t}$ has the correct engineering dimensions.

There is one major difference between condensed matter systems and
the early universe that is not addressed in equations like
(\ref{op}).  This is that, whereas the effective Ginzburg-Landau
field theories for condensed matter systems are complete, the
early universe contains many (probably most) fields about which we
know nothing, and which are ignored in ({\ref{op}). Only for QCD
to we have a full tally of the relevant degrees of freedom of the
system. The effect of this environment is to make the quantum
field theory behave classically (show decoherence)
\cite{lom1,lom2}. The easiest way to show this is by seeing how
the density matrix (whose diagonal elements are $p_t[\Phi ]$)
becomes diagonal as a consequence of the growth of the unstable
modes. In parallel, the master equation for the Wigner functional
plays the role of a Fokker-Planck equation, whose Langevin
counterpart for the semiclassical field is a variant of
(\ref{tdlg}), including multiplicative noise. However, none of
this interferes strongly with defining $t^*$ from (\ref{op}).

What is important is that, by the time that defects are produced
they behave like classical entities \cite{lom2}.

\subsection{\bf Other systems}

Not all systems behave as the above. Bose-Einstein condensates
satisfy the Gross-Pitaevski equation \cite{zhitnitsky}
\begin{equation}
i\hbar\frac{\partial\phi_{a}}{\partial t} = \frac{\delta
F}{\delta\phi_{a}}. \label{gp}
\end{equation}

Although causality still leads to the Zurek bounds, transitions
 do not evolve because of the
exponential growth of unstable modes. A similar situation exists
for those transitions that are driven by changes in chemical
potentials (as in high-$T_c$ superconductors).

\subsection{\bf Additional mechanisms for gauge theories}

All the mechanisms above are well-suited for global
symmetry-breaking, but for local symmetry-breaking in the presence
of gauge fields there is an additional mechanism for the
production of defects. This mechanism, observed by Hindmarsh and
Rajantie \cite{rajantie} is discussed in detail in Arttu
Rajantie's contribution to these proceedings, and will only be
considered briefly here.

For the most relevant case of superconductors (both high and low
temperature) the idea is simple, even if the execution is
difficult. Above the critical temperature there are thermal
fluctuations in the electromagnetic field. As the system cools
through $T_c$ the short wavelength modes of the field stay in
equilibrium, whereas long wavelength modes drop out of equilibrium
and freeze in.  These frozen modes are the source of correlated
flux that will be measured along with the flux of conventional
Abrikosov vortices produced from the causal arguments above. The
effect is not simply additive but will, in general, lead to a
greater variance in the spontaneously produced flux passing
through a surface.

\subsection{Defects as zeroes}

So far we have only been discussing the time $t^*$ it takes for
the transition to complete itself, and comparing this to the
causal time ${\bar t}$.  The important step is to derive the
defect separation $\xi^*$ at the time of defect production and
compare it to ${\bar \xi}$ of (\ref{xibar}).

To do this, we identify simple defects, with normal (false vacuum)
cores, by the zeroes of the order parameter fields. At early
times, when field fluctuations are strong, field zeroes do not
have the energy profiles to be identified with defects, whose
masses, tensions, etc, are $O(\eta)$ non-perturbatively large.
However, by $t^*$, the energy profiles are qualitatively correct,
and counting zeroes provides a reliable estimate of defect
numbers.
[In the case of QFT, this requires
that defects can be considered classically.]

In the linear regime the separation of zeroes is given \cite{maz}
simply in turns of the derivatives of the two-point correlation
function $\langle\phi_a({\vec x},t)\phi_b({\vec 0},t)\rangle$. In
this approximation we find that, just as $t^*\approx {\bar t}$ (up
to logarithmic corrections), then so is $\xi^*\approx {\bar\xi}$.

We have observed that many systems have defects with complicated
cores. Simple numerical simulations show that, provided the
non-normal components of the  cores are not too large, 
the scaling laws of the causal analysis are preserved
\cite{pedro1}.

 There is a further important
comment on thermal fluctuations. Initially \cite{kibble2}, it was
thought that domain size after a transition was determined by the
correlation length at the Ginzburg temperature $T_c$.
In the causal arguments above the Ginzburg
regime plays no role, but this reappears in the analysis of
unstable modes, where thermal fluctuations on small scales can
make the production of defects difficult \cite{ray2} if the quench
is too slow.

\section{Experimental Confirmation}

To date, several condensed matter experiments had been performed
to test (\ref{xibar}):
\begin{itemize}
\item {\bf Superfluid $^{3}He-B$}\cite{grenoble,helsinki}.
 The two experiments\cite{grenoble,helsinki} on $^{3}He-B$
rely on the fact that,
when superfluid $^{3}He-B$ is bombarded with slow neutrons, $%
n+^{3}He\rightarrow p+^{3}H+760$keV. The energy released in such a
collision leads to a hot spot  in the superfluid, with temperature
$T>T_{c}$, which when cooled by its environment, leaves behind a
tangle of vortices (the topological defects in this system). $\tau
_{Q}$ is fixed by the nuclear process that breaks up the $^{3}He$
atom. With only a single data point conflating both normalisation and $%
\sigma $ it is not possible to confirm the predicted value $\sigma
=1/4$. However, both experiments are highly compatible with
(\ref{xibar}).

\item {\bf Superfluid $^{4}He$}\cite{lancaster,lancaster2}. In
principle, the two $^{4}He$
experiments\cite{lancaster,lancaster2}, which use a pressure
quench,
allow for a more complete test.
Yet again, vortices are the relevant defects. In practice, the
most reliable experiment\cite{lancaster2} sees no vortices. This
is not necessarily a sign of contradiction in that it has been
suggested\cite{ray2} that the vortices decay too fast to be seen
in this case. This is irrespective\cite{ray} of whether high
thermal fluctuations within the wide Ginzburg regime of $^4He$
would lead to somewhat different predictions. In this context, the
vortices seen in an earlier $^{4}He$ experiment\cite {lancaster}
were most likely an artefact of the experimental setup.

\item {\bf High temperature superconductors
(HTSC)}\cite{technion,technion2}.  The two
experiments\cite{technion}, on HTSC measure the total flux of
Abrikosov vortices through a surface. The vortex separation of
(\ref{xibar}) can be converted into a prediction for the variance.
In the first experiment no flux was seen, despite the phase
separation that leads to the result being demonstrated
elsewhere\cite{carmi}. However, on increasing the quench rate by
several orders of magnitude spontaneous flux was produced
\cite{technion2}, but with low efficiency. The observed flux is compatible
with the Zurek prediction of (\ref{xibar}), once additional assumptions
for converting total flux into net flux are taken into account. Although the prediction (%
\ref{xibar}) has not taken gauge fields into account, the effect
of electromagnetic field fluctuations is expected to be
small\cite{rajantie2}.

\item {\bf Annular Josephson Tunnel Junctions}. The defects of a
lineaer JTJ are fluxons, the sine-Gordon model kinks in the field
$\phi =\phi _{1}-\phi _{2}$, the difference in the phases of the
complex order parameter fields in the separate superconductors.
Again two experiments have been performed (one currently in
progress, with data as yet unanalysed). The first experiment
\cite{roberto} was one in which, by counting fluxons on varying
quench time $\tau _{Q}$, it was possible to compare $\sigma $ with
its theoretical value (as well as confirming overall scale). The
secondary mechanism of \cite{rajantie} is not relevant here.
Agreement with both exponent and magnitude are good, although
there is scatter, with $\sigma = 0.27\pm 0.05$, in comparison to
the theoretical value of $\sigma = 0.25$. The new experiment
currently being performed will permit quench times four orders of
magnitude faster than before, and thereby an order of magnitude
more fluxons.

\item {\bf Other direct experiments}.  Two subsequent experiments
have permitted varying quench rates and
so an estimate for $\sigma $. The first \cite{pamplona} involves the B$%
\acute{e}$nard-Marangoni conduction-convection transition, in
which a homogeneous conduction state is broken into an hexagonal
array of convection lines on heating. The defects here are not
associated with the line zeroes of an order parameter field, and
the viscosity-dependent $\sigma $ does not match the ZK
prediction, possibly for that reason. The second
experiment\cite{florence} is carried out in a non-linear optical
system, with complex beam-phase the order parameter.
Increasing the
light intensity (the control parameter in this case) leads to
pattern formation (defects) at a critical value. The predicted
$\sigma =1/4$ is recovered to good accuracy as $\sigma
_{exp}=0.25\pm 0.02$.

\item {\bf Related experiments}. Related experiments include
measurements of defect production in continuous transitions of
planar liquid crystals \cite{digal,swarup}, measurements of flux
production in dilute Josephson Junction annuli \cite{kirtley}, and
measurements at first order transitions in radioactive low-$T_c$
superconductors \cite{girard}. The first provides a check on the
relationships between net and gross topological charge, the second
 on the role of electromagnetic field fluctuations
at a transition \cite{rajantie2}, relevant for the high-$T_c$
experiments. At the moment the implications of the third are
unclear, given the different nature of first order transitions.

\end{itemize}

\section{Conclusions}

  We have seen that the patterns of symmetry breaking in condensed matter systems and the early universe
  are sufficiently rich and sufficiently similar to pursue the
  comparison.

In almost all cases the transitions are signalled by the
production of defects, usually topological vortices, vortons or
monopoles. The observation of defects is therefore a demonstration
that the transitions have taken place. Defect production in the
early universe is complicated by the assumed inflationary period,
and there are only tantalising glimpses of possible candidates.
The situation is very different for condensed matter systems,
where defects are readily observed.

In this latter case we can do more, and examine the details of
defect production. A comparison of the experimental results with
the causal constraints on defect production proposed by Zurek
shows the strength and limitations of the causal bounds and
informs the similar arguments on the role of causality in defect
production in the early universe, as originally mooted by Kibble.

\section*{Acknowledgments}
We thank the organisers for their hospitality and the ESF for
financial support through the COSLAB programme.


\begin{thebibliography}{99}

\bibitem{COSLAB1} {\it Topological Defects and the Non-Equilibrium Dynamics of Symmetry Breaking Phase
Transitions}, Proceedings of the Nato ASI, Les Houches (2000),
Y.M. Bunkov and H. Godfrin (eds.), Nato Science Series {\bf C549},
Kluwer Academic Publishers (2001).


\bibitem{COSLAB2}{\it Patterns of Symmetry Breaking}, Proceedings
of the COSLAB meeting at Crakow (2002), H. Arodz et al. (eds.),
Kluwer Academic Publishers, Dordrecht (2003).

\bibitem{zhang} S-C. Zhang, eprint cond-mat/9610140, Science {\bf 275}, 1089 (1997)

\bibitem{arovas} D.P. Arovas, A.J. Berlinsky, C. Kallin and S-C. Zhang, {\it Phys. Rev.
Lett.}\, {\bf 79}, 2871 (1997).

\bibitem{alama} S. Alama, A.J. Berlinsky, L. Brossard and T.Giorgi,
 {\it Phys. Rev.}\, {\bf B60}, 6901 (1999).


\bibitem{volovik} {\it Exotic Properties of Superfluid $^3He$}, G.E.
Volovik, World Scientific, Singapore (1992).

\bibitem{Bunkov} Yu. M. Bunkov and O.D. Timofeevskaya, {\it Phys.
Rev. Lett.}\, {\bf 80}, 4927 (1998).

\bibitem{chuang} I. Chuang, R. Durrer, N. Turok and B. Yurke, {\it
Science}{\bf 251}, 1336 (1991)

\bibitem{digal} S. Digal. R.Ray and A.M. Srivastava {\it Phys.
Rev. Lett} {\bf 83}, 5030 (1999).

\bibitem{zhitnitsky} M.A. Metlitski and A.R. Zhitnitsky,
cond-mat/0307559 (2003)



\bibitem{rajagopal} K. Rajagopal, Chapter 35 in the Festschrift
in honor of B. L. Ioffe, "At the Frontier of Particle Physics /
Handbook of QCD", M. Shifman, ed., (World Scientific, Singapore),
2001, hep-ph/0011333.

\bibitem{korthes-althes}  C. Korthals-Altes, A. Kovner and M.
Stephanov, {\it Phys.Lett.} {\bf B469}, 205 (1999).

\bibitem{buckley} K.B.W. Buckley and A.R. Zhitnitsky, {\it
JHEP}{\bf 208}, 13 (2002)

\bibitem{witten} E. Witten, {\it Nucl. Phys} {\bf B249}, 557
(1985)

\bibitem{mairi} R. Jeannerot, J. Rocher and M. Sakellariadou
{\it Phys.Rev.} {\bf D68} 103514 (2003).

\bibitem{zhitnitsky2}  K.B.W. Buckley and A.R. Zhitnitsky, {\it
Phys. Rev.}{\bf B67} 174522 (2003).

\bibitem{sazhin} M. Sazhin et al.,
{\it Mon. Not. Roy. Astron. Soc.}{\bf 343},
353 (2003), astro-ph/0302547

\bibitem{bhattachargee} P. Bhattachargee and G. Sigl, {\it Phys.
Rep.}{\bf 327}, 109 (2000)

\bibitem{bonazzola} S. Bonazzola and P. Peter, {\it Astropart.
Phys.}{\bf 7}, 161 (1997)

\bibitem{boyanovsky2} D. Boyanovsky, H.J. de Vega, M. Simionato {\it The Early Universe and
the Cosmic Microwave Background: Theory and Observations}, N.G.
Sanchez and Y.N. Parijskij (eds.), Nato Science Series II {\bf
150}, Kluwer Academic Publishers (2003), 65.

\bibitem{glendenning} N.K. Glendenning and F. Weber, {\it
Astrophys. J.}{\bf480L} 111 (1997).



\bibitem{zurek1}  W.H. Zurek, {\it Nature} {\bf 317}, 505 (1985), {\it Acta
Physica Polonica} {\bf B24}, 1301 (1993).

\bibitem{zurek2}  W.H. Zurek, {\it Physics Reports} {\bf 276}, 177 (1996).

\bibitem{kibble1}  T.W.B. Kibble, in {\it Common Trends in Particle and
Condensed Matter Physics}, {\it Physics Reports} {\bf 67}, 183
(1980).






\bibitem{bettencourt} N. Antunes, L.M. Bettencourt, W. H. Zurek
 {\it Phys. Rev. Lett} {\bf 82}, 2824 (1999).


\bibitem{RKK} E. Kavoussanaki,
R.J. Rivers and G. Karra, {\it Condensed Matter Physics} {\bf 3},
133 (2000).

\bibitem{rayulti1} R.J. Rivers, {\it Journal of Low Temperature
Physics}\, {\bf 124}, 41-84, (2001).

\bibitem{boyanovsky} D. Boyanovsky, Da-Shin Lee and A. Singh,
{\it Phys. Rev.} {\bf D48}, 800 (1993), \\
D. Boyanovsky, H.J. de Vega and R. Holman, {\it Phys. Rev.} {\bf
D49}, 2769 (1994).

\bibitem{lom1}R.J. Rivers, F.C. Lombardo, F.D. Mazzitelli, {\it Phys. Lett.} {\bf B 523}, 317 (2002)

\bibitem{lom2}R.J. Rivers, F.C. Lombardo, F.D. Mazzitelli, {\it Phys. Lett.} {\bf B 539}, 1 (2002)



\bibitem{rajantie}  M. Hindmarsh and A.Rajantie, {\it Phys. Rev. Lett.}{\bf %
\thinspace 85}, 4660 (2000); A. Rajantie, {\it Journal of Low
Temperature Physics}{\bf \thinspace 124}, 5 (2001).


\bibitem{maz} F. Liu and G.F. Mazenko, {\it Phys. Rev.}\, {\bf
B46}, 5963 (1992).

\bibitem{pedro1} N. Antunes, P. Gandra, R.J. Rivers and A. Swarup,
in preparation.

\bibitem{kibble2} T.W.B. Kibble, {\it J. Phys.}{\bf A 9}, 1387
(1976).

\bibitem{ray2}  G. Karra and R.J. Rivers, {\it Phys. Rev. Lett.} {\bf 81},
3707 (1998).

\bibitem{grenoble}  C. Bauerle {\it et al.}, {\it Nature}\thinspace {\bf 382}%
, 332 (1996).

\bibitem{helsinki}  V.M.H. Ruutu {\it et al.}, {\it Nature}\thinspace {\bf %
382}, 334 (1996).



\bibitem{lancaster}  P.C. Hendry {\it et al}, {\it Nature}\thinspace {\bf 368%
}, 315 (1994).

\bibitem{lancaster2}  M.E. Dodd {\it et al.}, {\it Phys. Rev. Lett.}%
\thinspace {\bf 81}, 3703 (1998), {\it J. Low Temp. Physics}\thinspace {\bf %
15}, 89 (1999).

\bibitem{ray}  R.J. Rivers, {\it Phys. Rev. Lett} {\bf 84}, 1248 (2000).

\bibitem{technion}  R. Carmi and E. Polturak, {\it Phys. Rev.} {\bf %
B\thinspace 60}, 7595 (1999).



\bibitem{technion2} A. Maniv, E. Polturak and G. Koren, eprint
cond-mat/0304359.

\bibitem{carmi}  R. Carmi, E. Polturak, and G. Koren, {\it Phys. Rev. Letts.}
{\bf 84}, 4966 (2000).



\bibitem{rajantie2} T.W.B. Kibble and A. Rajantie, eprint
cond-mat/0306633.

\bibitem{roberto} R. Monaco, J. Mygind and R.J. Rivers {\it Phys. Rev. Lett.}
{\bf 89}, 080603 (2002); {\it Phys. Rev.} {\bf B 67}, 104506
(2003).

\bibitem{pamplona}  S. Casado, W. Gonz\'{a}lez-Vi\~{n}as, H. Mancini and S.
Boccaletti, {\it Phys. Rev.\thinspace }{\bf E63}, 057301 (2001).

\bibitem{florence}  S. Ducci, P.L. Ramazza, W. Gonz\'{a}lez-Vi\~{n}as, and
F.T. Arecchi, {\it Phys. Rev. Lett.\thinspace }{\bf 83}, 5210
(1999).

\bibitem{swarup} R.J. Rivers and A. Swarup, cond-mat/0312082.

\bibitem{kirtley} J.R. Kirtley, C.C. Tsuei and F. Tafuri,
arXiv:cond-mat/0304359

\bibitem{girard}J. Páramos, O. Bertolami, T.A. Girard, P. Valko, {\it Phys.Rev.} {\bf B67} 134511
(2003).



\end{thebibliography}
\end{document}